\documentclass[aps, pra, letterpaper, reprint, showpacs, longbibliography]{revtex4-1}

\pdfoutput=1

\usepackage{amsmath,amssymb,amsfonts,latexsym,color,dcolumn,bm,mathtools,amsbsy}
\usepackage[english]{babel}
\usepackage{graphicx}
\usepackage[utf8]{inputenc}
\usepackage{dsfont}
\usepackage{multirow}
\usepackage[per-mode=symbol]{siunitx}
\usepackage{textcomp}
\DeclareSIUnit{\sqrthz}{\ensuremath{\sqrt{\text{\hertz}}}}
\usepackage[babel=true,final=true,protrusion=true, expansion=true]{microtype}
\usepackage[colorlinks,urlcolor=blue,citecolor=blue]{hyperref}
\hypersetup{
	pdftitle={Measurement-induced long-distance entanglement of superconducting qubits using optomechanical transducers},
	pdfauthor={O. \v{C}ernot\'ik, K. Hammerer}
}

\newcommand{\D}{\ensuremath{\mathcal{D}}}

\renewcommand{\H}{\ensuremath{\mathcal{H}}}
\renewcommand{\L}{\ensuremath{\mathcal{L}}}
\newcommand{\dd}{\ensuremath{{d}}}
\newcommand{\dt}{\dd{}t}
\newcommand{\dW}{\dd{}W}
\newcommand{\Hint}{\ensuremath{H_\mathrm{int}}}
\newcommand{\avg}[1]{\ensuremath{\langle #1\rangle}}
\newcommand{\Hc}{\ensuremath{\mathrm{H.c.}}}

\newcommand{\covU}{\ensuremath{\Gamma^\mathrm{u}}}
\newcommand{\covC}{\ensuremath{\Gamma^\mathrm{c}}}

\newcommand{\nbar}{\ensuremath{\bar{n}}}

\newcommand{\ket}[1]{\ensuremath{|#1\rangle}}
\newcommand{\bra}[1]{\ensuremath{\langle #1|}}
\newcommand{\eff}{\ensuremath{\mathrm{eff}}}
\newcommand{\meas}{\ensuremath{\mathrm{meas}}}
\newcommand{\mech}{\ensuremath{\mathrm{mech}}}
\newcommand{\zpf}{\ensuremath{\mathrm{zpf}}}
\newcommand{\om}{\ensuremath{\omega_{m}}}
\newcommand{\chim}{\ensuremath{\chi_{m}}}
\newcommand{\q}{\ensuremath{{q}}}
\newcommand{\T}{\ensuremath{{T}}}
\newcommand{\rmi}{i}
\newcommand{\rme}{e}
\newcommand{\tr}{\ensuremath{\mathrm{tr}}}

\begin{document}

\title[Entanglement of superconducting qubits using optomechanical transducers]
    {Measurement-induced long-distance entanglement of superconducting qubits using optomechanical transducers}

\author{Ond\v{r}ej \v{C}ernot\'ik}
\email{Ondrej.Cernotik@itp.uni-hannover.de}
\affiliation{Institut f\"ur Theoretische Physik, Institut f\"ur Gravitationsphysik (Albert-Einstein-Institut),
    Leibniz Universit\"at Hannover, Callinstra\ss{}e 38, 30167 Hannover, Germany}

\author{Klemens Hammerer}
\affiliation{Institut f\"ur Theoretische Physik, Institut f\"ur Gravitationsphysik (Albert-Einstein-Institut),
    Leibniz Universit\"at Hannover, Callinstra\ss{}e 38, 30167 Hannover, Germany}

\date{\today}

\begin{abstract}
Although superconducting systems provide a promising platform for quantum computing, their networking poses a challenge
    as they cannot be interfaced to light---the medium used to send quantum signals through channels at room temperature.
We show that mechanical oscillators can mediated such coupling and light can be used to measure the joint state of two distant qubits.
The measurement provides information on the total spin of the two qubits such that entangled qubit states can be postselected.
Entanglement generation is possible without ground-state cooling of the mechanical oscillators
    for systems with optomechanical cooperativity moderately larger than unity;
    in addition, our setup tolerates a substantial transmission loss.
The approach is scalable to generation of multipartite entanglement
    and represents a crucial step towards quantum networks with superconducting circuits.
\end{abstract}


\maketitle

\section{Introduction}

Superconducting systems \cite{Blais2007, Arakawa2015} are among the best candidates for future quantum computers,
    owing to on-chip integration, scalability, precise control, and strong nonlinearities.
Recent experiments demonstrated their fast initialization \cite{Johnson2012, Riste2012a, Riste2012},
    observation and stabilization of their quantum trajectories \cite{Vijay2012, Murch2013, DeLange2014, Weber2014, Tan2015},
    quantum error correction \cite{Corcoles2015, Kelly2015, Riste2015}, and entanglement generation
    through parity measurements \cite{Riste2013, Saira2014, Roch2014}.

Scalable approaches towards quantum computers and networks \cite{Kimble2008} based on superconducting systems will require
    links between superconducting qubits to bridge long distances through room-temperature environments
    while preserving coherence and entanglement.
Such quantum-coherent links, based on light, have been demonstrated for single atoms, atomic ensembles, or artificial atoms
    (such as solid-state impurities or quantum dots) \cite{Hammerer2010,Sangouard2011,Reiserer2015,Lodahl2015}.
Roch et al. have made the first step towards similar networks with superconducting systems by entangling two qubits in separate cavities
    connected by a $\SI{1.3}{\metre}$  low-loss electrical wire at cryogenic temperature \cite{Roch2014,Motzoi2015,Chantasri2016}.
Extending this approach to room-temperature channels poses a significant challenge owing to lack of coupling to light
    which can transmit quantum information over long distances.
    
\begin{figure}[b]
	\centering
	\includegraphics[width=0.8\linewidth]{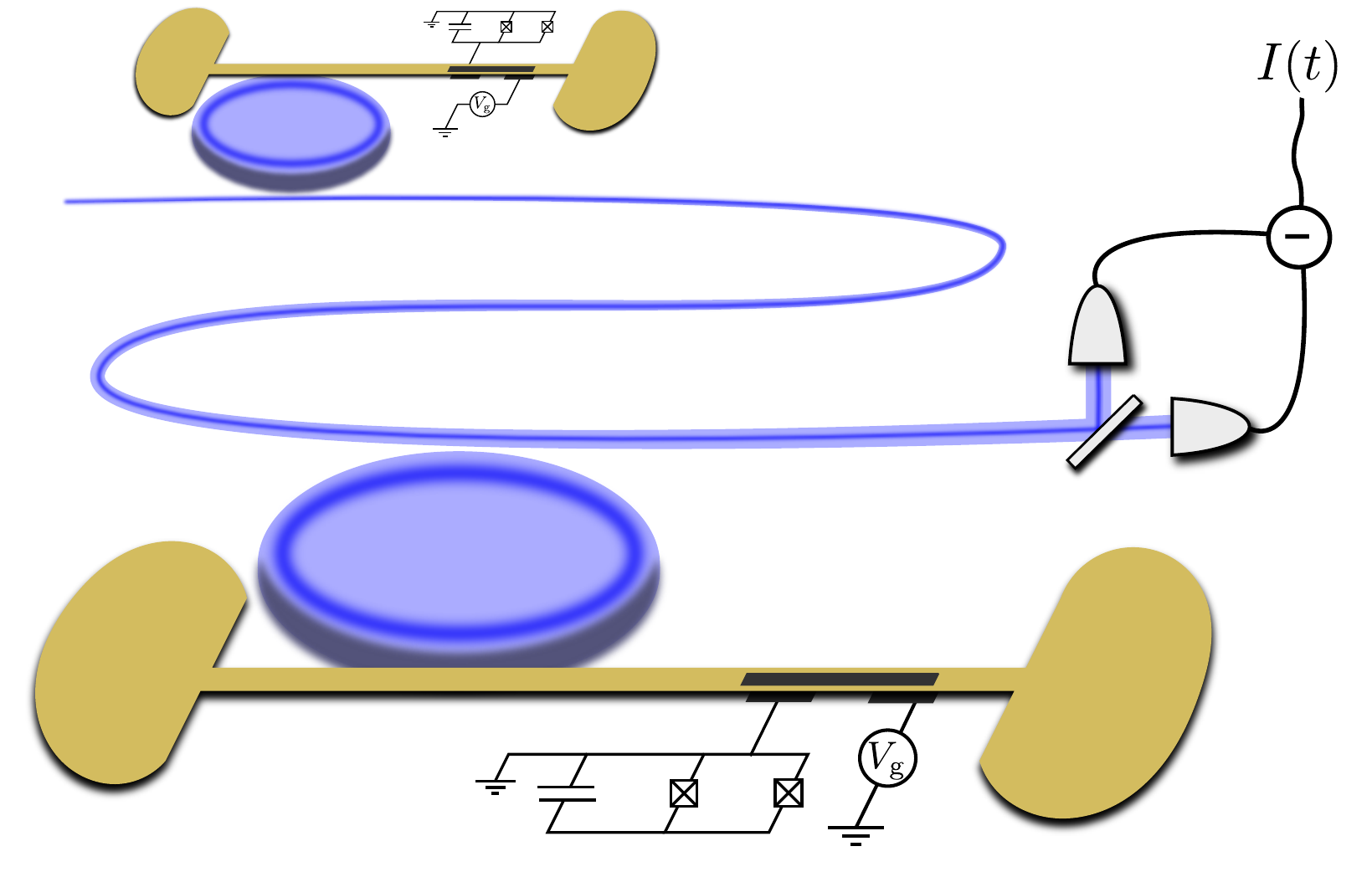}
	\caption{\label{fig.parity}(Color online)
		Schematic of the setup.
		Each of two qubits (e.g., superconducting transmon qubits, shown as black circuits) interacts with a mechanical oscillator
            (nanobeam, shown in yellow) that, in turn, couples to an optical mode.
        The optical resonators (blue toroids) are unidirectionally coupled; the output of the first (top) cavity enters the second (bottom) one.
		Homodyne measurement at the output of the second cavity provides information about the joint state of the two qubits 
            and can be used to postselect an entangled state.}
\end{figure}

Various auxiliary systems have been suggested as solutions for this issue
    by mediating the interaction between superconducting qubits and light.
One approach relies on atomic, molecular, or solid-state impurity spins, exploiting their magnetic and optical dipole moments 
    \cite{Tian2004,Sorensen2004,Rabl2006,OBrien2014, Blum2015, Xia2015};
    spin waves in ferromagnetic materials can use similar principles in coupling superconducting systems to light
    \cite{Tabuchi2015,Tabuchi2015a,Hisatomi2016}.
Recently, mechanical oscillators attracted attention as transducers between microwave circuits and light
    \cite{Taylor2011, Barzanjeh2012, Tian2012, Wang2012},
    and great experimental progress has been reported in this direction \cite{Bochmann2013, Andrews2014, Bagci2014}.
Optomechanical transducers benefit from the versatile interactions of mechanical oscillators with light through radiation pressure
    \cite{Aspelmeyer2013} and with microwave circuits via electrostatic \cite{Armour2002, Cleland2004, Regal2008, Heikkila2014}
    or magnetic forces \cite{Xue2007, Xia2014}.
The progress towards optomechanical transducers is paralleled by advances in coupling superconducting qubits
    with mechanical oscillators \cite{Etaki2008,LaHaye2009, Pirkkalainen2013, Pirkkalainen2014,Rouxinol2016} but it remains a challenge
    to identify simple, efficient schemes for integrating superconducting qubits and light in a single, hybrid system.
Theoretical proposals so far \cite{Clader2014, Yin2015,Stannigel2010} considered sophisticated time-dependent protocols
    that involve complex control schemes
    and require an unprecedented coupling strength (corresponding to optomechanical cooperativities of several hundred
    in the example studied in Ref. \cite{Yin2015} and up to thousands in the proposal by Stannigel et al. \cite{Stannigel2010})
    between mechanical oscillators and electromagnetic fields.

We demonstrate that entangled states of superconducting qubits can be generated by continuous homodyne detection of light.
Our scheme, shown in Fig.~\ref{fig.parity}, works in analogy to the technique established by Roch et al.
    in the microwave domain \cite{Roch2014} but employs an optical channel at room temperature.
Since our strategy builds on established experimental techniques, it is straightforward, though challenging,
    to implement it in a practical realization.
The use of light instead of microwaves as a link between the qubits greatly enhances the distance over which the qubits can become entangled;
    our work thus presents a crucial step towards quantum networks of superconducting circuits.
The generalization of the experiment of Roch et al. to the optical domain is, however, highly nontrivial and requires
    a systematic investigation of new sources of decoherence---thermal mechanical noise and optical transmission losses
    have to be addressed.
Such an analysis is complicated since the Hilbert space dimensions involved are too large,
    prohibiting even numerical Monte Carlo simulations;
    an alternative approach, based on adiabatic elimination of the complex transducer dynamics \cite{Cernotik2015}, is necessary.

Compared to earlier proposals, our strategy requires no time-dependent control. 
This simplicity leads to modest requirements on the system parameters; optomechanical cooperativity moderately larger than unity suffices,
    presenting an improvement of two to three orders of magnitude from previous proposals \cite{Stannigel2010,Yin2015}.
Thanks to resonant driving, sideband resolution is not necessary; furthermore the transducer acts as a force sensor
    in which light is used to detect the state-dependent force the qubit exerts on the mechanical oscillator.
The scheme is surprisingly resilient to photon losses---transmittance of 20 \% suffices for entanglement generation in the case study detailed below.
Finally, our theoretical treatment is completely general, admitting a broad range of experimental implementations---including
    superconducting transmon or flux qubits and magnetic sublevels of nitrogen-vacancy centers \cite{Wrachtrup2006}---with current technology.

We present the main idea in Sec.~\ref{sec:results} and discuss possible experimental platforms in Sec.~\ref{sec:implementations}.
In the second half of the paper, we discuss technical details of the scheme:
We derive effective equations of motion for the qubits in Sec.~\ref{sec:elimination}
    and discuss numerical methods in Sec.~\ref{sec:numerics}.
Finally, we conclude in Sec.~\ref{sec:conclusions}.

\section{Results}\label{sec:results}

\subsection{Optomechanical transducer as a force sensor}

First, we analyze a single node of the system shown in Fig.~\ref{fig.parity}.
A single qubit interacts dispersively with a mechanical oscillator (as shown, e.g., in Refs. \cite{Pirkkalainen2013,Pirkkalainen2014})
    via the Hamiltonian $\Hint = -Fx\sigma_z= \hbar\chi(b+b^\dagger)\sigma_z$.
The oscillator [with position operator $x= \sqrt{2}x_\zpf(b+b^\dagger)$] feels a force with magnitude
    $F= \hbar\chi/\sqrt{2}x_\zpf$ and direction that depends on the state of the qubit;
    conversely, the qubit experiences a frequency shift $\chi$ per displacement in zero-point fluctuation $x_\zpf = \sqrt{\hbar/2m_\eff\om}$ 
    ($m_\eff$ is the effective mass and $\om$ the angular frequency of the oscillator).
The position of the mechanical oscillator is then measured optically using a resonantly driven cavity field:
In an optomechanical measurement, the mechanical displacement determines a phase shift on light as described by the Hamiltonian
    $H_\mathrm{om} = \hbar g(a+a^\dagger)(b+b^\dagger)$ ($a$ denotes the annihilation operator of the optical field)
    \cite{Aspelmeyer2013}.
Thus, we can infer the force from homodyne detection of the phase quadrature of the cavity output.
The (shot-noise limited) sensitivity of an optomechanical force sensor at Fourier frequency $\omega$ is given
    by the spectral density of added force noise $S^2_F(\omega)=\kappa x^2_\zpf/[8g^2\chim^2(\omega)]$
    where $\kappa$ is the cavity linewidth and $\chim(\omega)=\left[m_\eff(\om^2-\omega^2)-\rmi m_\eff\gamma\omega\right]^{-1}$
    is the mechanical susceptibility \cite{Clerk2010, Aspelmeyer2013} ($\gamma$ is the width of the mechanical resonance).
The forces $\pm F$ corresponding to the qubit states $\ket{0}$, $\ket{1}$ can be distinguished by measuring for time
\begin{equation}\label{eq.sensitivity}
    \tau_\meas = \frac{S_F^2(\omega)}{F^2} = \frac{\kappa\om^2}{16\chi^2 g^2}.
\end{equation}
In the last step, we took into account that the qubit exerts a quasi-static force
    and the optomechanical force measurement concerns frequencies $\omega\ll\om$ and $\chim(\omega) \simeq 1/m_\eff\om^2$.
The measurement time $\tau_\meas$ needs to be shorter than the lifetime of the qubits whose decoherence we divide in two parts:
    the intrinsic lifetime (characterized by the relaxation and dephasing lifetimes $T_{1,2}$)
    and decoherence due to the interaction with the transducer.

To compare the measurement time with the intrinsic lifetime, we consider a transmon qubit with $T_{1,2}\approx\SI{20}{\micro\second}$.
For qubit-mechanical coupling $\chi = 2\pi\times\SI{5}{\mega\hertz}$ \cite{Pirkkalainen2013, Pirkkalainen2014},
    the required force sensitivity $S_F = F\sqrt{\tau_\meas}\sim\SI{0.5}{\femto\newton\per\sqrthz}$,
    which is a challenging but achievable precision.
We can estimate the mechanically induced decoherence from the force the oscillator exerts on the qubit, 
    $f(\omega) = \chi x(\omega)/\sqrt{2}x_\zpf$ with $x(\omega)= \chim(\omega) f_\mathrm{th}(\omega)$;
    here, $f_\mathrm{th}(\omega)$ with spectrum $S_\mathrm{th}^2(\omega) = 2\gamma m_\eff \hbar\om\nbar$ is the thermal force
    acting on the mechanical oscillator and $\nbar\simeq k_\mathrm{B}T/\hbar\om$ is the mean occupation of the oscillator
    in thermal equilibrium at temperature $T$.
The mechanical-dephasing rate of the qubit is given by the spectral density of the force $f$ at $\omega\ll\om$,
\begin{equation}\label{eq.GammaMec}
    \Gamma_\mech=S_f^2(\omega) = \frac{2\chi^2}{\om^2}\gamma\nbar.
\end{equation}
Comparing the dephasing time $\tau_\mathrm{deph} = 1/\Gamma_\mech$
    (associated with the mechanically induced dephasing) with the measurement time $\tau_\meas$,
    we see that the measurement dominates for large optomechanical cooperativity, 
    $C = 4g^2/\kappa\gamma\nbar>\frac{1}{2}$;
    in this regime, thermal noise does not limit the measurement of the qubit state.
Yet, the cooperativity cannot be too large since measurement backaction would start to disturb the qubit state
    and this simple model would break down.

\subsection{Generation of entanglement by measurement}

With an intuitive understanding of a single system, we now analyse two nodes connected via a unidirectional optical link,
    cf. Fig.~\ref{fig.parity}.
In this setup, the homodyne current reveals the total spin of the qubits $\sigma_z^1+\sigma_z^2$.  
The states $\ket{01}$, $\ket{10}$ give rise to the same measurement signal; if we prepare the qubits in the state
    $\ket{+}\ket{+}$ with $\ket{+} = (\ket{0}+\ket{1})/\sqrt{2}$ and the measurement outcome is $\avg{\sigma_z^1+\sigma_z^2} = 0$,
    the qubits end in the entangled state $\ket{\Psi_+} = (\ket{01}+\ket{10})/\sqrt{2}$
    \cite{Roch2014,Hutchison2009,Motzoi2015,Chantasri2016}.
We expect that entanglement can be prepared efficiently under the conditions discussed above, $T_{1,2}>\tau_\meas$ and $C>\frac{1}{2}$.

We check this expectation rigorously using a conditional master equation \cite{Wiseman2010,Zhang2014}
    that describes the evolution of the state $\rho(t)$ of the entire system conditioned on the homodyne measurement current $I(t)$
    (we put $\hbar=1$ in the following)
\begin{align}\label{eq.SME}
    \dd\rho &= -\rmi[H, \rho]\dt+\mathcal{L}_\q\rho\dt+\sum_{j=1}^2\gamma\{(\nbar+1)\D[b_j]+\nbar\D[b_j^\dagger]\}\rho\dt \nonumber\\
        &\quad+ \kappa\D[a_1-a_2]\rho\dt + \sqrt{\kappa}\H[(a_1-a_2)\rme^{\rmi\phi}]\rho\dW,\\
    I\dt &= \sqrt{\kappa}\avg{(a_1-a_2)\rme^{\rmi\phi}+\Hc}\dt+\dW.\label{eq:current}
\end{align}
The Hamiltonian $H =H_\mathrm{int} + H_\mathrm{mech}+H_\mathrm{om}+H_\mathrm{casc}$,
    where $H_\mathrm{int} = \chi[(b_1+b_1^\dagger)\sigma_z^1+(b_2+b_2^\dagger)\sigma_z^2]$ accounts for coupling of the qubits
    to the mechanical oscillators, $H_\mathrm{mech} = \om(b_1^\dagger b_1+b_2^\dagger b_2)$ is the free mechanical Hamiltonian,
    $H_\mathrm{om} = g[(a_1+a_1^\dagger)(b_1+b_1^\dagger)+(a_2+a_2^\dagger)(b_2+b_2^\dagger)]$ gives the optomechanical coupling,
    and $H_\mathrm{casc} = \rmi\kappa(a_1a_2^\dagger -a_2a_1^\dagger )/2$---together with the Lindblad term
    $\kappa\D[a_1-a_2]\rho$---describes the cascaded cavity dynamics \cite{Gardiner2004}.
Furthermore, $\mathcal{L}_\q\rho$ denotes intrinsic qubit relaxation and dephasing with lifetimes $T_{1,2}$,
    the Lindblad terms are given by
    $\D[O]\rho = O\rho O^\dagger - \frac{1}{2}(O^\dagger O\rho+\rho O^\dagger O)$, and the last term,
    with $\H[O]\rho = (O-\avg{O})\rho + \rho(O^\dagger-\avg{O^\dagger})$, accounts for the effect of continuous homodyne measurement;
    $\dW$ is Wiener increment with mean value $E(\dW) = 0$ and variance $E(\dW^2) = \dt$.
Finally, in Eq. \eqref{eq:current}, $\phi$ is the local oscillator phase.

Eq.~\eqref{eq.SME} cannot be integrated directly, either analytically or numerically.
Even simulations of quantum trajectories would need to include prohibitively large Hilbert space dimensions
    owing to the thermal occupation numbers (tens to hundreds in the examples below) of the two mechanical oscillators.
Instead, we adiabatically eliminate the linear, Gaussian dynamics of the transducer from Eq. \eqref{eq.SME}
    \cite{Cernotik2015}; see Sec.~\ref{sec:elimination} for technical details.
Provided the qubit coupling $\chi$ is slow on the timescale of the dynamics of the transducer
    (dominated, in the examples below, by the optical decay rate $\kappa$),
    we obtain an effective equation for the reduced state $\rho_\q$ of the two qubits,
\begin{equation}\label{eq.SME_qubits}
    \dd\rho_\q =\mathcal{L}_\q\rho_\q\dt+ \L_1\rho_\q\dt + \L_2\rho_\q\dt + \sqrt{\Gamma_\meas}\H[\sigma_z^1+\sigma_z^2]\rho_\q\dW,
\end{equation}
    which can be treated both numerically and analytically and which reveals the role of individual parameters of the transducer.
Here, $\L_1\rho_\q = \Gamma_\mech\sum_j\D[\sigma_z^j]\rho_\q$ describes mechanically induced dephasing of the qubits
    and $\L_2\rho_\q = \Gamma_\meas\D[\sigma_z^1+\sigma_z^2]\rho_\q$ accounts for the measurement backaction.
The measurement and dephasing rates are given by 
\begin{equation}\label{eq.rates}
    \Gamma_\meas = 16\frac{\chi^2g^2}{\kappa\om^2},\qquad\Gamma_\mech = \frac{\chi^2\gamma}{\om^2}(2\nbar+1),
\end{equation}
    in perfect agreement with the simple argument of force sensing (with $\Gamma_\meas = \tau_\meas^{-1}$);
    cf. Eqs.~\eqref{eq.sensitivity}, \eqref{eq.GammaMec}.
We assume that the measurement phase $\phi$ is properly optimized (in the limit of weak optomechanical coupling,
    this optimization corresponds to detection of the phase quadrature, $\phi = \pi/2$).
Finally, the measurement current provides information on the total spin of the qubits,
\begin{equation}
    I\dt = 2\sqrt{\Gamma_\meas}\avg{\sigma_z^1+\sigma_z^2}\dt+\dW.
\end{equation}

Last but not least, we can also include optical losses.
Photon loss between the two systems affects only the first qubit and leads to its dephasing (like mechanical noise does); 
    losses after the second system influence both qubits equally and limit the detection efficiency.
Although limited detection efficiency does not introduce additional decoherence of the qubits,
    it necessitates longer measurement times for which other decoherence processes will degrade the state.
We can include these effects, as well as any asymmetry in system parameters, in the master equation \eqref{eq.SME_qubits};
    see Sec.~\ref{sec:elimination} for details.

\begin{table*}
	\caption{\label{tab.parameters}
		Suggested experimental parameters for the systems shown in Fig.~\ref{fig.implementations}.
		We consider transmon qubits (Sec.~\ref{ssec:transmon}) coupled to either nanobeams or vibrating membranes,
            flux qubits (Sec.~\ref{ssec:flux}), and nitrogen-vacancy centers (Sec.~\ref{ssec:NV}).
		We assume the systems are cooled to \SI{20}{\milli\kelvin} temperature.}
	\begin{tabular}{@{}llllll}
	   \hline\hline
		 &&\multicolumn{2}{c}{Transmon}&&\\
		 \cline{3-4}
		Quantity & Symbol & Nanobeam & Membrane & Flux qubit & NV center \\
		\hline
		Qubit coupling (kHz) & $\chi/2\pi$ & 5800 & 3700 & 2400 & 50 \\
		Mechanical frequency (MHz) & $\om/2\pi$ & 8.7 & 1.0 & 2.3 & 2.0 \\
		Effective mass (pg) & $m_\eff$ & 3.8 & 30 000 & 10 & 25 \\
		Mechanical quality & $Q_m$ & $5\times 10^4$ & $5\times 10^5$ & $10^5$ & $10^4$ \\
		Mechanical decay (Hz) & $\gamma/2\pi$ & 170 & 2 & 22 & 200 \\
		Thermal occupation & $\nbar$ & 48 & 420 & 185 & 210 \\
		Optical quality & $Q_\mathrm{opt}$ & $5\times 10^6$ & $10^7$ & $10^7$ & $2\times 10^8$ \\
		Optical decay (MHz) & $\kappa/2\pi$ & 39 & 19 & 19 & 1 \\
		Bare optomechanical coupling (Hz) & $g_0/2\pi$ & 300 & 25 & 65 & 20 \\
		Optomechanical coupling (kHz) & $g/2\pi$ & 900 & 140 & 450 & 300 \\
		Driving power (\si{\micro\watt}) & $P$ & 138 & 300 & 370 & 90 \\
		Optomechanical cooperativity & $C$ & 10 & 5 & 10 & 8.5 \\
		Measurement rate (kHz) & $\Gamma_\mathrm{meas}/2\pi$ & 150 & 230 & 190 & 0.9 \\
		Force sensitivity (\si{\atto\newton\per\sqrthz}) & $S_F$ & 130 & 1900 & 38 & 18 \\
		Displacement sensitivity (\si{\atto\metre\per\sqrthz}) & $S_x$ & 11 & 1.6 & 18 & 4.1 \\
		Schematic figure && \ref{fig.implementations}(a) & \ref{fig.implementations}(b) & \ref{fig.implementations}(c) 
            & \ref{fig.implementations}(d) \\
		\hline\hline
	\end{tabular}
\end{table*}

\section{Implementations}\label{sec:implementations}

\begin{figure}
	\centering
	\includegraphics[width=\linewidth]{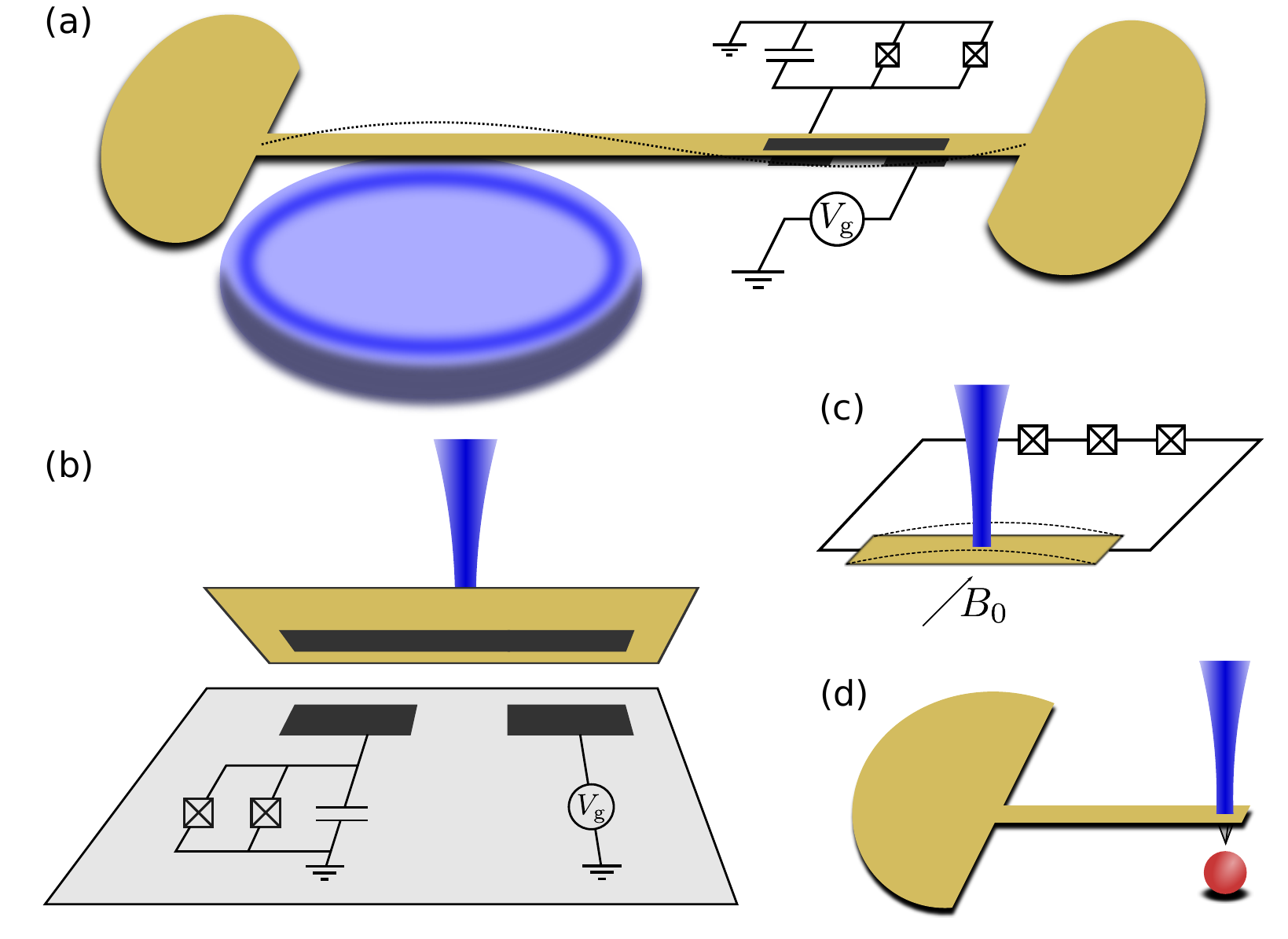}
	\caption{\label{fig.implementations}(Color online)
		Possible experimental realizations of the proposed scheme.
		Transmon qubits can interact with mechanical oscillators via position-dependent gate capacitance.
        The mechanical oscillator can be formed by a nanobeam that interacts with the near field
            of a toroidal optical resonator (a), or a membrane used as an end mirror or in the middle of a Fabry-Perot cavity (b).
		Mechanical oscillators can further be integrated into the circuit of a flux qubit (c).
		Other solid-state qubits---such as magnetic sublevels of a nitrogen-vacancy centers---can be used as well,
            exploiting magnetic coupling to cantilevers (d).
		Superconducting circuits are indicated in black, nitrogen-vacancy center in red, mechanical oscillators in yellow,
            and optical modes in blue.
	}
\end{figure}

The discussion so far was completely general and did not assume any specific realization. 
In this section, we discuss several possible implementations as shown in Fig.~\ref{fig.implementations}:
The most promising realization relies on superconducting transmon qubits coupled to mechanical motion
    using a mechanically compliant gate capacitor \cite{Heikkila2014, Pirkkalainen2013}.
For this setup, we will integrate the effective master equation \eqref{eq.SME_qubits}
    and show that entanglement can indeed be generated under the conditions derived above.
Later, we briefly treat two further setups in which these requirements can be met:
First, flux qubits with mechanical oscillators forming a part of the qubit loop \cite{Xue2007} and, second, solid-state-spin qubits---such
    as magnetic levels of nitrogen-vacancy centers---that can
    interact with mechanical oscillators via magnetic fields \cite{Rabl2009, Rabl2010, Hong2012a, Kolkowitz2012}.
We list experimental parameters of these systems in Tab.~\ref{tab.parameters}.

\subsection{Transmon qubits}\label{ssec:transmon}

The first implementation uses superconducting transmon qubits 
    that interact with mechanical oscillators via mechanically compliant gate capacitors $C_{g} = C_{g}(x)$
    \cite{Armour2002, Heikkila2014,Pirkkalainen2013}.
We describe the qubit using its canonical operators---the charge $n$ and the phase $\varphi$ across the qubit island---and
    write the total Hamiltonian as \cite{Pirkkalainen2013}
\begin{align}\label{eq:transmon}
    H &= 4E_C(n-n_0)^2 - E_J(\Phi)\cos\varphi-E_-(\Phi)\sin\varphi \nonumber\\
    &\quad+\omega_m b^\dagger b + \chi(n_0-n)(b+b^\dagger).
\end{align}
Here, $E_C$ denotes charging energy, $n_0$ charge induced by gate voltage $V_g$,
    and we express the Josephson energy using the sum and difference of the Josephson energies of the two junctions $E_{J1,2}$;
    the Josephson energy is controlled using an external flux $\Phi$,
\begin{subequations}
\begin{align}
    E_J(\Phi) &= (E_{J1}+E_{J2})\cos\left(\pi\frac{\Phi}{\Phi_0}\right),\\
    E_-(\Phi) &= (E_{J1}-E_{J2})\sin\left(\pi\frac{\Phi}{\Phi_0}\right).
\end{align}
\end{subequations}
    Finally, the coupling rate for a parallel-plate capacitor \cite{Armour2002}
\begin{equation}
    \chi = 2E_{C}\frac{C_{g}V_{g}}{e}\frac{x_\mathrm{zpf}}{d}
\end{equation}
    where $e$ is the elementary charge and $d$ the distance between the capacitor and the mechanical oscillator.

Considering only the first two levels of the transmon generally gives
    a free Hamiltonian with a non-negligible transversal component,
    which after diagonalization translates into transversal coupling between the qubit and the mechanical oscillator,
    $\chi_x(b+b^\dagger)\sigma_x$.
Nevertheless, for flux $\Phi = \Phi_0/2$ and equal Josephson energies $E_{J1}=E_{J2}$, 
    the corresponding terms in the Hamiltonian \eqref{eq:transmon} are identically zero
    and the qubit Hamiltonian has only the longitudinal component,
\begin{align}
    H = \frac{\omega_q}{2}+\omega_m b^\dagger b+\chi(b+b^\dagger)\sigma_z.    
\end{align}
Small discrepancies between the Josephson energies do not pose a problem;
    the resulting transversal coupling is weak and can be neglected in the rotating-wave approximation.

The mechanical oscillator can take the form of a nanobeam which interacts with the evanescent field of a microtoroidal cavity 
    \cite{Anetsberger2009, Wilson2014}, see Fig.~\ref{fig.implementations}(a).
In such a system, the opto- and electromechanical parts of the system are well spatially separated
    and photon absorption will not heat the superconducting circuit.
Alternatively, we can use a vibrating membrane forming an end mirror or placed in the middle of a Fabry-Perot cavity;
    such designs have been used in recent experimental demonstrations of microwave-to-optical conversion 
    \cite{Andrews2014, Bagci2014}, cf. Fig.~\ref{fig.implementations}(b).
The system of Andrews et al. \cite{Andrews2014} is particularly relevant, since it uses the second harmonic mode
    of a membrane-in-the-middle setup; the optical and microwave field interact with different antinodes of the motion,
    which minimizes optical heating of the superconducting circuit.

\subsubsection{Nanobeam mechanical oscillators}

We consider a silicon nitride beam of length $l = \SI{70}{\micro\metre}$ that is $w = \SI{400}{\nano\metre}$ wide
    and $t = \SI{100}{\nano\metre}$ thick.
The gate capacitor and the microtoroid both lie at an antinode of the second harmonic mode of frequency
    $\om/2\pi \approx \SI{8.7}{\mega\hertz}$ with effective mass $m_\mathrm{eff} = \SI{3.8}{\pico\gram}$
    and zero-point motion $x_\mathrm{zpf} = \SI{16}{\femto\metre}$.
If one third of the beam is covered by a superconductor,
    the gate capacitance $C_{g} = \SI{275}{\atto\farad}$ for a beam-circuit distance $d_{q} = \SI{75}{\nano\metre}$.
For a typical charging energy $E_{C} = \SI{5}{\giga\hertz}$ and gate voltage $V_{g} = \SI{10}{\volt}$,
    the qubit coupling $\chi/2\pi \approx \SI{5.8}{\mega\hertz}$.
Note that coupling rate comparable with the mechanical frequency has recently been demonstrated \cite{Pirkkalainen2014}.

For the optomechanical system, we consider a beam-toroid distance of $d_\mathrm{om} = \SI{50}{\nano\metre}$ and a toroid with radius $r = \SI{30}{\micro\metre}$,
    resulting in bare optomechanical coupling of about $g_0/2\pi \approx \SI{300}{\hertz}$.
Driving the optical resonator with the power 
\begin{equation}
    P = \frac{1}{2}\hbar \frac{2\pi c}{\lambda}\kappa \left(\frac{g}{g_0}\right)^2 \approx \SI{138}{\micro\watt}
\end{equation}
    for linewidth $\kappa/2\pi = \SI{39}{\mega\hertz}$
    (toroidal whispering gallery resonators can have decay rates an order of magnitude smaller \cite{Anetsberger2009}),
    we can reach optomechanical coupling $g/2\pi \approx \SI{900}{\kilo\hertz}$
    and a measurement rate $\Gamma_\mathrm{meas}/2\pi \approx \SI{150}{\kilo\hertz}$,
    corresponding to a measurement time of about \SI{1}{\micro\second}.
Such a measurement is strong enough to be performed within the coherence time of transmon qubits which is typically around
    \SIrange{10}{20}{\micro\second} \cite{Riste2013, Roch2014}.

\subsubsection{Entanglement generation with nanobeam oscillators}

\begin{figure}
	\centering
	\includegraphics[width=\linewidth]{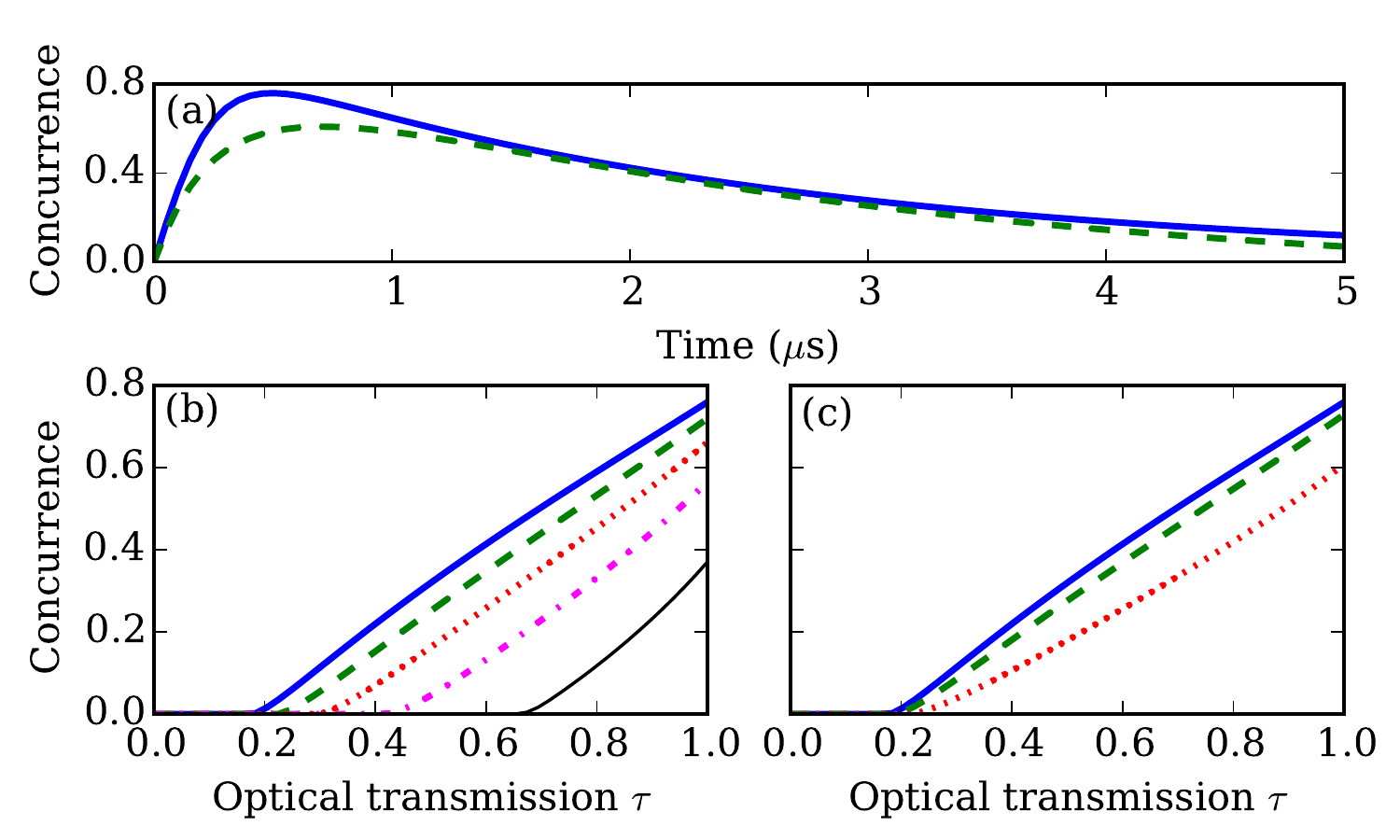}
	\caption{\label{fig.concurrence}(Color online)
		Entanglement generation with the nanobeam optomechanical transducer of Fig.~\ref{fig.implementations}(a).
		In panel (a), we show the time dependence of the concurrence for success probability
            $P_\mathrm{succ} = 0.1$ (solid blue line) and $P_\mathrm{succ} = 0.5$ (dashed green line).
		In the bottom panels, we plot the concurrence (optimized over the measurement time) versus optical transmission $\tau$
            between the two nodes.
        In (b), we plot the concurrence for several detection efficiencies [$\eta = 1$ (solid blue line),
            $\eta = 0.8$ (dashed green line), $\eta = 0.6$ (dotted red line), $\eta = 0.4$ (dot-dashed magenta line),
            and $\eta = 0.2$ (thin black line)]; in (c), the plotted curves represent different success probabilities
            [$P_\mathrm{succ} = 0.1$ (solid blue line), $P_\mathrm{succ} = 0.3$ (dashed green line),
            and $P_\mathrm{succ} = 0.5$ (dotted red line)].
		We consider qubit coupling $\chi = 2\pi\times\SI{5.8}{\mega\hertz}$,
            mechanical frequency $\om = 2\pi\times\SI{8.7}{\mega\hertz}$ and quality $Q_\mathrm{m} = 5\times 10^4$,
            optical decay rate $\kappa = 2\pi\times\SI{39}{\mega\hertz}$, optomechanical coupling $g = 2\pi\times\SI{900}{\kilo\hertz}$,
            and temperature \SI{20}{\milli\kelvin} (corresponding to thermal occupation $\nbar = 48$);
            we assume the intrinsic relaxation and coherence lifetimes of the qubits $T_{1,2} = \SI{20}{\micro\second}$.
		Furthermore, we use the values $\tau = \eta = 1$ for panel (a), $P_\mathrm{succ} = 0.1$ for (b), and $\eta = 1$ for (c).}
\end{figure}

We study entanglement generation with the nanobeam mechanical oscillators in Fig.~\ref{fig.concurrence}.
To this end, we consider the following protocol for entanglement generation:
Measuring for time $T$, we accumulate the total signal $J(T) = \int\dt\,I(t)$.
We then compare this signal with a predefined postselection cutoff $\nu$
and keep the state if and only if $|J(T)|<\nu$.
Two parameters are of interest to us---the entanglement of the resulting state and the success probability, i.e.,
the probability of the signal being small than the cutoff.
We discuss the protocol in more detail in Sec.~\ref{sec:numerics} and derive a simplified model
that enables us to find the resulting state analytically without generating quantum trajectories.
In Fig.~\ref{fig.concurrence}(a), we plot the concurrence \cite{Wootters1998} of the final state as a function of time.
At early times, $t<\tau_\meas/2$, the measurement is inconclusive owing to overlap of signals corresponding to different outcomes,
    resulting in strongly mixed postselected states.
Next, the concurrence reaches its maximum around $t\sim \tau_\meas$ and then steadily decays because of dephasing and relaxation of the qubits.
In the following, we thus use the optimum value as a figure of merit characterizing the scheme.

In Fig.~\ref{fig.concurrence}(b), we analyze how optical losses affect the concurrence.
We investigate transmission losses between the two nodes (horizontal axis) as well as finite detection efficiency
    (individual curves in the plot, see the figure caption for more details).
Remarkably, entanglement can be generated with up to 80 \% loss;
    it is possible to generate entanglement in presence of higher losses
    if the qubit lifetimes and optomechanical cooperativity are increased.
Finally, in panel (c), we study transmission losses in combination with success probability.
While larger success probability generally leads to a smaller concurrence,
    it has little effect on the transmission loss for which the concurrence reaches zero.

\subsubsection{Oscillating membranes}

Next, we consider a membrane-in-the-middle optomechanical system,
    similar to Ref. \cite{Andrews2014} with membrane dimensions \SI{1}{\milli\metre}$\times$\SI{1}{\milli\metre}
    and second harmonic frequency of $\om/2\pi = \SI{1}{\mega\hertz}$ placed $d_{q} = \SI{500}{\nano\metre}$
    from the gate capacitor with capacitance $C_{g} = \SI{60}{\femto\farad}$.
In such a system, qubit-mechanical coupling $\chi/2\pi = \SI{3.7}{\mega\hertz}$ can be achieved.
With a modest optomechanical system with coupling $g/2\pi = \SI{140}{\kilo\hertz}$ and decay rate $\kappa/2\pi = \SI{19}{\mega\hertz}$,
    the effective measurement rate can reach value of \SI{230}{\kilo\hertz}.


\subsection{Flux qubits}\label{ssec:flux}

For measurements with flux qubits, we consider the system proposed in Ref. \cite{Xue2007},
    see Fig.~\ref{fig.implementations}(c).
A mechanical oscillator forms a part of the qubit loop and the persistent current through the loop $I_p$
    together with an external magnetic field $B_0$ act with a qubit-state-dependent Lorentz force on the oscillator.
The coupling rate is given by
\begin{equation}
    \chi = B_0 I_{p} l_\mathrm{eff} x_\mathrm{zpf},
\end{equation}
    where $l_\mathrm{eff}$ is the effective length of the mechanical oscillator.
For a \SI{12}{\micro\metre} long bridge, the mechanical frequency $\om/2\pi = \SI{2.3}{\mega\hertz}$,
    effective mass $m_\eff = \SI{10}{\pico\gram}$, and coupling rate $\chi/2\pi = \SI{2.3}{\mega\hertz}$.
The bridge can form one end of a Fabry-Perot cavity;
    with coupling rate $g/2\pi = \SI{450}{\kilo\hertz}$ and optical decay rate $\kappa/2\pi = \SI{19}{\mega\hertz}$, 
    the effective measurement rate $\Gamma_\mathrm{meas}/2\pi = \SI{190}{\kilo\hertz}$.
Since flux qubits have shorter lifetimes than transmons (typically around \SI{5}{\micro\second}),
    weaker entanglement can be generated in such a system.
Moreover, integration of the mechanical oscillator into the superconducting circuit
    as well as the optomechanical Fabry-Perot cavity will lead to absorption heating of the circuit.

\subsection{Nitrogen-vacancy centers}\label{ssec:NV}

The protocol is not limited to superconducting systems.
Here, we study entanglement of the magnetic sublevels of electron spins in nitrogen-vacancy centers
    using the transducer schematically depicted in Fig.~\ref{fig.implementations}(d).
The qubit interacts with a cantilever that has a magnetic tip
    and that serves as an end mirror of a Fabry-Perot optical cavity.
In such a system, magnetomechanical coupling $\chi/2\pi = \SI{50}{\kilo\hertz}$ can be expected \cite{Rabl2009,Stannigel2010}.
For a mechanical frequency $\om/2\pi = \SI{2}{\mega\hertz}$, optomechanical coupling $g/2\pi = \SI{300}{\kilo\hertz}$,
    and optical linewidth $\kappa/2\pi = \SI{1}{\mega\hertz}$ (thus requiring extremely high quality Fabry-Perot resonator),
    the effective measurement rate is about $\Gamma_\mathrm{meas}/2\pi = \SI{0.9}{\kilo\hertz}$,
    requiring qubit lifetime on the order of milliseconds;
    such values of the magnetomechanical coupling, cavity decay rate, and dephasing lifetime put substantial requirements
    on the fabrication of the system.

Alternatively, other kinds of solid-state spins can be used.
For instance, the coherence lifetime of phosphorus donors in silicon can reach several seconds \cite{Tyryshkin2012};
    with such systems, we can relax the requirements on the magnetomechanical coupling and the optical decay.
If we consider the values $\chi/2\pi = \SI{10}{\kilo\hertz}$, $\kappa/2\pi = \SI{10}{\mega\hertz}$ (with other parameters same as before), 
    the qubit measurement rate $\Gamma_\meas/2\pi = \SI{3}{\hertz}$, corresponding to a measurement time of about \SI{50}{\milli\second}.

\section{Derivation of effective equations of motion}\label{sec:elimination}

In this section, we derive the effective master equation \eqref{eq.SME_qubits} that describes the dynamics of the qubits.
We start with a single node and show how we can adiabatically eliminate the mechanical and optical degrees of freedom.
Then, we discuss two such systems connected with a directional optical link;
    after treating an idealized system, we analyze the role of various imperfections,
    specifically, the presence of optical loss and asymmetry in the parameters of the two transducers.

\subsection{Single-qubit readout}

We start from the stochastic master equation for a qubit coupled to a single transducer (i.e., a single optomechanical system),
\begin{align}\label{eq.single_node}
    \dd\rho &= -\rmi[H,\rho]\dt + \L_\q\rho\dt + \gamma\{(\nbar+1)\D[b]+\nbar\D[b^\dagger]\}\rho\dt \nonumber\\
        &\quad+ \kappa\D[a]\rho\dt + \sqrt{\kappa}\H[a\rme^{\rmi\phi}]\rho\dW
\end{align}
    with the Hamiltonian $H = \chi(b+b^\dagger)\sigma_z + \om b^\dagger b + g(a+a^\dagger)(b+b^\dagger)$.
To adiabatically eliminate the mechanical and optical degrees of freedom, we consider the equation of motion of the transducer
\begin{align}
    \dd\rho_\T &= -\rmi[\om b^\dagger b + g(a+a^\dagger)(b+b^\dagger), \rho_\T]\dt \nonumber\\
        &\quad+ \gamma\{(\nbar+1)\D[b]+\nbar\D[b^\dagger]\}\rho_\T\dt + \kappa\D[a]\rho_\T\dt \nonumber\\
        &\quad+ \sqrt{\kappa}\H[a\rme^{\rmi\phi}]\rho_\T\dW,
\end{align}
    where the subscript T indicates that the density matrix $\rho_\T$ describes the state of the transducer.
Since the dynamics is linear, the state $\rho_T$ is fully described by the first and second statistical moments of the canonical operators,
    which we collect in the vector
\begin{equation}
    r = \frac{1}{\sqrt{2}} [a+a^\dagger, -\rmi(a-a^\dagger), b+b^\dagger, -\rmi(b-b^\dagger)]^T;
\end{equation}
    we are particularly interested in the covariance matrix with elements
\begin{align}
    \Gamma_{ij} &= \avg{[r_i-\avg{r_i},r_j-\avg{r_j}]_+} \nonumber\\
    &= \tr\{[r_i-\avg{r_i},r_j-\avg{r_j}]_+\rho_\T \}.
\end{align}
The covariance matrix of the conditional state of the transducer obeys the Riccati equation \cite{Cernotik2015}
\begin{subequations}
\begin{align}
    \dot{\Gamma}^\mathrm{c} &= A\Gamma^\mathrm{c}+\Gamma^\mathrm{c} A^T + 2N \nonumber\\
        &\quad-2(\Gamma^\mathrm{c} c-\sigma m)(\Gamma^\mathrm{c} c-\sigma m)^T,\label{eq.Riccati}\\
    A &= \left(\begin{array}{cccc} -\kappa/2&0&0&0 \\ 0&-\kappa/2&-2g&0 \\ 0&0&-\gamma/2&\om \\ -2g&0&-\om&-\gamma/2 \end{array}\right), \\
    N &= \mathrm{diag}\left[\frac{\kappa}{2}, \frac{\kappa}{2}, \gamma\left(\nbar+\frac{1}{2}\right),
        \gamma\left(\nbar+\frac{1}{2}\right)\right], \\
    c &= \frac{\kappa}{\sqrt{2}}(\cos\phi, -\sin\phi, 0,0)^T,\\
    m &= \frac{\kappa}{\sqrt{2}}(\sin\phi, \cos\phi, 0,0)^T,\\
    \sigma_{ij} &= -\rmi[r_i,r_j];
\end{align}
\end{subequations}
on the other hand, the unconditional state of the transducer follows the deterministic master equation
\begin{align}
    \dot{\rho}_\T^\mathrm{u} &= -\rmi[\om b^\dagger b + g(a+a^\dagger)(b+b^\dagger), \rho_\T^\mathrm{u}] \nonumber\\
        &\quad+ \gamma\{(\nbar+1)\D[b]+\nbar\D[b^\dagger]\}\rho_\T^\mathrm{u} + \kappa\D[a]\rho_\T^\mathrm{u}
\end{align}
    and the corresponding covariance matrix obeys the Lyapunov equation
\begin{equation}\label{eq.Lyapunov}
    \dot{\Gamma}^\mathrm{u} = A\Gamma^\mathrm{u} + \Gamma^\mathrm{u} A^T+2N.
\end{equation}
We use the superscripts c, u to distinguish the covariance matrix of the conditional and unconditional state.

The effective stochastic master equation for the qubit is (we assume summation over $i$, $j$, and $k$) \cite{Cernotik2015}
\begin{subequations}
\begin{align}
    \dd\rho_\q &= \L_\q\rho_\q\dt + \frac{1}{2}A_{ij}^{-1}\covU_{jk}[s_i,[s_k,\rho_\q]]\dt\nonumber\\
        &\quad+\frac{\rmi}{2}A_{ij}^{-1}\sigma_{jk}[s_i,[s_k,\rho_\q]_+]\dt \nonumber\\
        &\quad+ \H[\rmi\Lambda^Ts]\rho_\q\dW,\label{eq.single_qubit} \\ 
    \Lambda &= (\covC-\rmi\sigma)Q^{-T}c+A^{-1}(\covC c-\sigma m), \\ 
    Q &= A-2(\covC c-\sigma m)c^T,\\ 
    s &= (0,0,\sqrt{2}\chi,0)^T.
\end{align}
\end{subequations}
We thus need to solve the Riccati and Lyapunov equations \eqref{eq.Riccati}, \eqref{eq.Lyapunov};
    these equations can be solved analytically in the limit of weak optomechanical coupling, $g<\kappa$,
    which corresponds to shot-noise limited readout, but we omit the resulting expressions for brevity.
Plugging everything in, we find the effective equation
\begin{align}
    \dd\rho_\q &= \L_\q\rho_\q\dt + (\Gamma_\mathrm{meas}+\Gamma_\mathrm{mech})\D[\sigma_z]\rho_\q\dt \nonumber\\
        &\quad+ \sqrt{\Gamma_\mathrm{meas}}\H[\sigma_z]\rho_\q\dW,
\end{align}
    where the measurement and dephasing rates $\Gamma_\meas$, $\Gamma_\mech$ are given in Eq.~\eqref{eq.rates}.

\subsection{Two-qubit measurement}

The dynamics of two cascaded systems are described by the equation
\begin{align}\label{eq.two-qubit}
    \dd\rho &= -\rmi[H, \rho]\dt+ \L_\q\rho\dt + \kappa\D[a_1-a_2]\rho\dt \nonumber\\
        &\quad+  \gamma\{(\nbar+1)(\D[b_1]+\D[b_2])+\nbar(\D[b_1^\dagger]+\D[b_2^\dagger])\}\rho\dt\nonumber\\
        &\quad	 + \sqrt{\kappa}\H[(a_1-a_2)\rme^{\rmi\phi}]\rho\dW
\end{align}
    with the Hamiltonian
\begin{align}\label{eq:Hamiltonian}
    H &= \sum_{j=1}^2 [\chi\sigma_z^j(b_j+b_j^\dagger) + \omega_m b_j^\dagger b_j + g(a_j+a_j^\dagger)(b_j+b_j^\dagger)] 
        \nonumber\\
        &\quad+ \rmi\frac{\kappa}{2}(a_1a_2^\dagger-a_2a_1^\dagger).
\end{align}
Apart from local dynamics of the two nodes---given by the square bracket in the Hamiltonian \eqref{eq:Hamiltonian}
    and by the dissipation of the qubits and the mechanical oscillators---there is the unidirectional coupling of the optical cavities.
This effect appears in the last term of the Hamiltonian \eqref{eq:Hamiltonian}
    and in the Lindblad and measurement terms $\D[a_1-a_2]\rho$,
    $\H[(a_1-a_2)\rme^{\rmi\phi}]\rho$.
(The minus sign in these terms is due to the choice of the relative phase between the two cavity fields,
    see the discussion below.)
We assume that the two qubit-oscillator-cavity systems are characterized by the same frequencies, coupling constants,
    and decoherence rates.

Adiabatic elimination of the mechanical and optical degrees of freedom can be done in complete analogy with the single-qubit readout
    and the effective two-qubit equation is
\begin{align}\label{eq.two_qubit}
	\dd\rho_\q &= \L_\q\rho_\q\dt + \Gamma_\mathrm{mech}\{\D[\sigma_z^1]+\D[\sigma_z^2]\}\rho_\q\dt \nonumber\\
	   &\quad + \Gamma_\mathrm{meas}\D[\sigma_z^1+\sigma_z^2]\rho_\q\dt \nonumber\\
	   &\quad + \sqrt{\Gamma_\mathrm{meas}}\H[\sigma_z^1+\sigma_z^2]\rho_\q\dW.
\end{align}
The relative phase between the two qubits in the measurement and two-qubit dephasing terms is set by the phase
    between the two cavities and can be controlled by applying an additional phase shift to the light field between the cavities.
The relevant choices are $\sigma_z^1+\sigma_z^2$
    [which can be used to generate the entangled state $\ket{\Psi_+} = (\ket{01}+\ket{10})/\sqrt{2}$]
    and $\sigma_z^1-\sigma_z^2$ [with which the state $\ket{\Phi_+} = (\ket{00}+\ket{11})/\sqrt{2}$ can be prepared];
    we can obtain the latter measurement from the former by applying a $\pi$ shift between the cavities, $a_2\to -a_2$.
Any other phase results in the signals from the two qubits appearing in different quadratures.

With optical losses in the system, the overall dynamics are described by the equation
\begin{align} 
	\dd\rho &= -\rmi[H,\rho]\dt + \L_\q\rho\dt + \kappa_1(1-\tau)\D[a_1]\rho\dt \nonumber\\
        &\quad + \sum_{j=1}^2\gamma_j\{(\nbar_j+1)\D[b_j]+\nbar_j\D[b_j^\dagger]\}\rho\dt \nonumber\\
        &\quad	 + \D[\sqrt{\kappa_1\tau}a_1-\sqrt{\kappa_2}a_2]\rho\dt \nonumber\\
        &\quad+ \sqrt{\eta}\H[(\sqrt{\kappa_1\tau}a_1 -\sqrt{\kappa_2}a_2)\rme^{\rmi\phi}]\rho\dW
\end{align}
    where the Hamiltonian
\begin{align}
    H &= \sum_{j=1}^2[\chi_j\sigma_z^j(b_j+b_j^\dagger) + \omega_{{m},j}b_j^\dagger b_j + g_j(a_j+a_j^\dagger)(b_j+b_j^\dagger)] 
        \nonumber\\
        &\quad- \frac{\rmi}{2}\sqrt{\kappa_1\kappa_2\tau}(a_1^\dagger a_2-a_2^\dagger a_1).
\end{align}
Here, $\tau\in(0,1]$ is the transmittance of the channel between the two cavities (including optical losses in the first cavity)
    and $\eta\in(0,1]$ is the detection efficiency (it includes any optical losses in and after the second cavity);
    moreover, we now assume different parameters for the two qubits, mechanical oscillators, and optical cavities.
The effective two-qubit equation of motion becomes
\begin{widetext}
\begin{align}
    \dd\rho_\q &= \L_\q\rho_\q\dt + \left(16(1-\tau)\frac{\chi_1^2g_1^2}{\kappa_1\omega_{{m},1}^2}
        +\frac{\chi_1^2\gamma_1(2\nbar_1+1)}{\omega_{{m},1}^2}\right)\D[\sigma_z^1]\rho_\q\dt
        + \frac{\chi_2^2\gamma_2(2\nbar_2+1)}{\omega_{{m},2}^2}\D[\sigma_z^2]\rho\dt  \nonumber\\
    &\quad + \D\left[\sqrt{16\tau\frac{\chi_1^2g_1^2}{\kappa_1\omega_{{m},1}^2}}\sigma_z^1
        +\sqrt{16\frac{\chi_2^2g_2^2}{\kappa_2\omega_{{m},2}^2}}\sigma_z^2\right]\rho_\q\dt
        +\sqrt{\eta}\H\left[\sqrt{16\tau\frac{\chi_1^2g_1^2}{\kappa_1\omega_{{m},1}^2}}\sigma_z^1
        +\sqrt{16\frac{\chi_2^2g_2^2}{\kappa_2\omega_{{m},2}^2}}\sigma_z^2\right]\rho_\q\dW.
\end{align}
\end{widetext}
For a total-spin measurement, we require that both qubits be measured at the same rate,
    $\tau\chi_1^2g_1^2/\kappa_1\omega_{{m},1}^2 = \chi_2^2g_2^2/\kappa_2\omega_{{m},2}^2$;
in most implementations, we can tune one of the couplings
    (the qubit coupling is usually tuneable using external fields, similar to the optomechanical coupling).
Here, we consider tuning the coupling of the second qubit, $\chi_2 = \sqrt{\tau}\chi_1$,
    since enhancing any of the coupling strengths in the first node would increase the dephasing rate of the first qubit.
The resulting equation of motion is then
\begin{widetext}
\begin{align}\label{eq.effective}
	\dd\rho_\q &=  \sum_{j=1}^2 \left(\frac{1}{T_1}\D[\sigma_-^j]+\frac{1}{T_2}\D[\sigma_z^j]\right)\rho_\q\dt 
	   +\left(16(1-\tau)\frac{\chi^2g^2}{\kappa\om^2}+\frac{\chi^2\gamma(2\nbar+1)}{\om^2}\right)\D[\sigma_z^1]\rho_\q\dt \nonumber\\
	&\quad + \frac{\tau\chi^2\gamma(2\nbar+1)}{\om^2}\D[\sigma_z^2]\rho\dt 
	   + 16\tau\frac{\chi^2g^2}{\kappa\om^2}\D\left[\sigma_z^1+\sigma_z^2\right]\rho_\q\dt  
	   + \sqrt{16\tau\eta\frac{\chi^2g^2}{\kappa\om^2}}\H\left[\sigma_z^1+\sigma_z^2\right]\rho_\q\dW \nonumber\\
	&= \sum_{j=1}^2 \left(\frac{1}{T_1}\D[\sigma_-^j]+\frac{1}{T_2}\D[\sigma_z^j]\right)\rho_\q\dt 
	   + [(1-\tau)\Gamma_\meas+\Gamma_\mech]\D[\sigma_z^1]\rho_\q\dt \nonumber\\
	&\quad+ \tau\Gamma_\mech\D[\sigma_z^2]\rho_\q\dt + \tau\Gamma_\meas\D[\sigma_z^1+\sigma_z^2]\rho_\q\dt 
	   + \sqrt{\tau\eta\Gamma_\meas}\H[\sigma_z^1+\sigma_z^2]\rho_\q\dW,
\end{align}
\end{widetext}
    where the mechanical coupling rate of the first and second qubit is $\chi$, $\sqrt{\tau}\chi$
    and $\Gamma_\meas$, $\Gamma_\mech$ are given in Eq.~\eqref{eq.rates};
    moreover, we wrote the intrinsic qubit decoherence explicitly
    using relaxation and dephasing processes with corresponding lifetimes $T_{1,2}$.
Individual tuning of the coupling rates (both $\chi$ and $g$) can also be used to compensate other differences
    between the systems arising during manufacture, such as the difference in optical decay rates or mechanical frequencies.

\section{Numerical methods}\label{sec:numerics}

Here, we discuss details of the postselection procedure and our approach to approximating the resulting state.
We start from the conditional master equation, which we write as
\begin{subequations}
\begin{align}
    \dd\rho &= \gamma_-\{\D[\sigma_-^1] + \D[\sigma_-^2]\}\rho\dt + \gamma_1\D[\sigma_z^1]\rho\dt \nonumber\\
    &\quad + \gamma_2\D[\sigma_z^2]\rho\dt + \Gamma\D[\sigma_z^1+\sigma_z^2]\rho\dt \nonumber\\
    &\quad+ \sqrt{\eta\Gamma}\H[\sigma_z^1+\sigma_z^2]\rho\dW,\\ 
    I\dt &= 2\sqrt{\eta\Gamma}\avg{\sigma_z^1+\sigma_z^2}\dt+\dW. 
\end{align}
\end{subequations}
We assume that both qubits relax at the same rate whereas their dephasing rates differ.
(This situation describes two identical qubits coupled to light via optomechanical transducers with optical losses between them.)
Now, we prepare the qubits in the state $\ket{\psi_0} = (\ket{0}+\ket{1})(\ket{0}+\ket{1})/2$
    and measure for time $T$, accumulating the signal
\begin{equation}\label{eq.signal}
    J(T) = \int_0^T I\dt.
\end{equation}
If $J(T) \approx 0$, the expectation value $\avg{\sigma_z^1+\sigma_z^2} = 0$
    and the qubits are in the entangled state $\ket{\Psi_+} = (\ket{01}+\ket{10})/\sqrt{2}$
    (assuming all decoherence channels are negligible compared with the measurement)
    whereas for $J(T)\ll 0$ they are in the state $\ket{11}$ [$\ket{00}$ for $J(T)\gg 0$].
Choosing a postselection cutoff $\nu$, we keep the state if $|J(T)|\leq\nu$ and discard it otherwise.

To get a deeper understanding of the dynamics, we adopt the following simplified approach:
We assume that the system first evolves according to the \emph{unconditional} master equation
\begin{align}\label{eq.deterministic}
    \dot{\rho} &= \gamma_-\{\D[\sigma_-^1] + \D[\sigma_-^2]\}\rho + \gamma_1\D[\sigma_z^1]\rho + \gamma_2\D[\sigma_z^2]\rho \nonumber\\
    &\quad+ \Gamma\D[\sigma_z^1+\sigma_z^2]\rho
\end{align}
    from time $t = 0$ to time $t = T$.
Afterwards, we perform a fast, strong measurement, which returns the result $J(T)$.
Finally, using the cutoff $\nu$, we either keep or discard the state; we ask how strong the entanglement in the final state is.
This approach is generally not valid since nonlinearity in the measurement term
    $\H[\sigma_z^1+\sigma_z^2]\rho = (\sigma_z^1+\sigma_z^2)\rho - \tr\{(\sigma_z^1+\sigma_z^2) \rho \}\rho+\Hc$
    mixes the three subspaces corresponding to the measurement outcomes,
    $\avg{\sigma_z^1+\sigma_z^2} = 2$ (spanned by the state $\ket{00}$), $\avg{\sigma_z^1+\sigma_z^2} = 0$ (spanned by $\ket{01}$,
    $\ket{10}$), and $\avg{\sigma_z^1+\sigma_z^2} = -2$ (spanned by $\ket{11}$),
    which are independent in the unconditional dynamics (assuming weak relaxation of the qubits).
Nevertheless, if the measurement is strong enough (so that the inter-subspace coherences quickly decay),
    this treatment well approximates the true stochastic dynamics which can otherwise be studied only using quantum trajectories.

Formally, we solve the deterministic master equation Eq.~\eqref{eq.deterministic}
    with the initial condition $\rho(t=0) = \ket{\psi_0}\bra{\psi_0}$;
    although this equation can be solved analytically, we omit the solution for brevity. 
The qubits then interact with the measurement apparatus (initially in the vacuum state),
    which we project on an eigenstate of the measurement operator and obtain the unnormalized state
\begin{equation}
    \tilde{\rho}_x = \bra{x}\exp(-\rmi\mu S_zp)\rho(T)\otimes\ket{0}\bra{0}\exp(\rmi\mu S_zp)\ket{x}.
\end{equation}
Here, $S_z = (\sigma_z^1+\sigma_z^2)/2$, $p$ is the phase quadrature of the measurement apparatus,
    and $\mu$ is the measurement strength, which we evaluate from the classical signal in Eq.~\eqref{eq.signal}:
Each of the projections $\avg{S_z} = 0, \pm 1$ gives rise to normally distributed signals $J(T)$
    with mean value $4\sqrt{\eta\Gamma}\avg{S_z}T$ and variance $T$.
The measurement apparatus is a bosonic mode and its interaction with the two-qubit system leads to displacement of this mode;
    the measurement strength is given by the mean (for $\avg{S_z} = 1$) renormalized by the square root of the variance and
    by factors coming from the definition of the amplitude quadrature $x = (a+a^\dagger)/\sqrt{2}$, so that $\mu = 2\sqrt{\eta\Gamma T}$.

The unnormalized projected state $\tilde{\rho}_x$ can be expressed using phase-quadrature representation for the measurement apparatus
\begin{align}
    \tilde{\rho}_x &= \frac{1}{(2\pi)^{3/2}} \int\dd p \exp\left(-\frac{p^2}{4} + \rmi(x-\mu S_z)p\right) \rho(T) \nonumber\\
        &\quad\times\int\dd p' \exp\left(-\frac{p'^2}{4} -\rmi(x-\mu S_z)p'\right)\nonumber\\
    &= \sqrt{\frac{2}{\pi}} \sum_{S,S' = -1}^1 \rme^{-(x-\mu S)^2} \mathbb{P}_S \rho(T) \mathbb{P}_{S'} \rme^{-(x-\mu S')^2}\nonumber\\
    &= D(x)\rho(T)D(x);
\end{align}
here, $\mathbb{P}_S$ is the projector onto the subspace with $\avg{S_z} = S$ and 
\begin{align}
    D(x) &= \sqrt[4]{\frac{2}{\pi}}\,\mathrm{diag}\left[e^{-(x-\mu)^2}, e^{-x^2}, e^{-x^2}, e^{-(x+\mu)^2}\right].
\end{align}
Integrating over the interval $x\in(-\nu,\nu)$, we obtain the postselected state
\begin{equation}
    \rho_f = \frac{\int_{-\nu}^{\nu}\dd x \tilde{\rho}_x}{\tr\{\int_{-\nu}^{\nu}\dd x \tilde{\rho}_x\}};
\end{equation}
    the normalization factor gives the success probability of the postselection procedure,
    $P_\mathrm{succ} = \tr\{\int_{-\nu}^{\nu}\dd x \tilde{\rho}_x\}$.
Although it is possible to express the final density matrix analytically,
    the resulting expression is too cumbersome to be presented here.

\begin{figure}
	\centering
	\includegraphics[width=\linewidth]{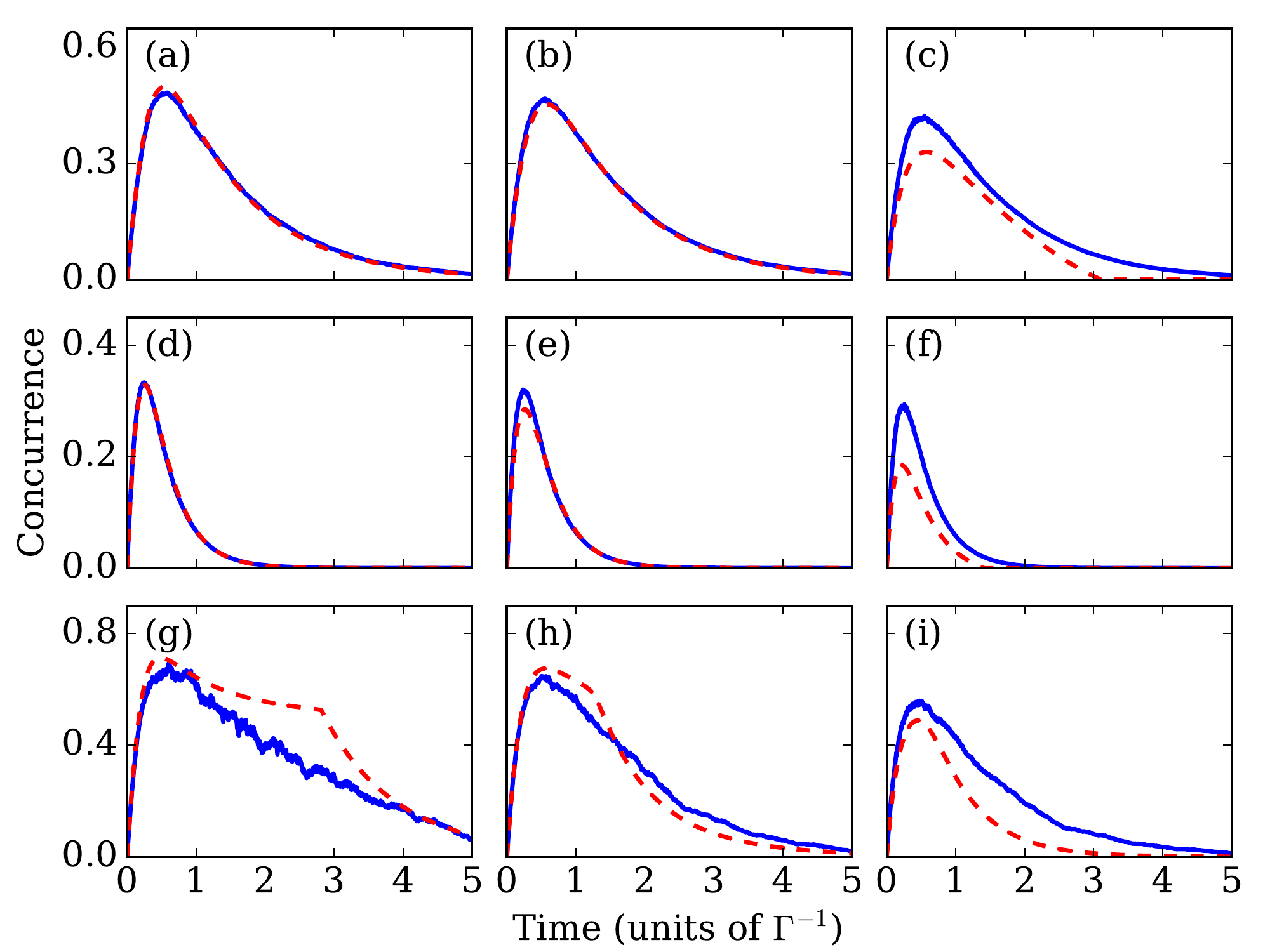}
	\caption{\label{fig.simulations}(Color online)
		Comparison of the analytical model (dashed red line) with numerical simulations (solid blue line).
		In the first row [panels (a)-(c)], we consider system parameters $\gamma_- = 0.1\Gamma$, $\gamma_1 = \gamma_2 = 0.2\Gamma$, 
            $\eta = 0.6$, in the second row [panels (d)-(f)], the parameters are $\gamma_- = 0.1\Gamma$, $\gamma_1=\Gamma$,
            $\gamma_2 = 0.3\Gamma$, $\eta = 1$, and for the last row [panels (g)-(i)], we use the parameters $\gamma_- = 0.8\Gamma$,
            $\gamma_1 = \gamma_2 = 0$, $\eta = 1$.
		The success probability $P_\mathrm{succ} = 0.1$ for the first column [panels (a), (d), (g)], $P_\mathrm{succ} = 0.3$
            for the second column [panels (b), (e), (h)], and $P_\mathrm{succ} = 0.5$ for the last column [panels (c), (f), (i)].}
\end{figure}

We compare the analytical model with numerical simulations in Fig.~\ref{fig.simulations}.
As expected, the analytical model breaks down when the relaxation rate becomes large
    and the nonlinearity of the measurement term starts to play a role [panels (g)-(i)].
Furthermore, the analytical model and the numerical simulations start to deviate for higher success probability [panels (c), (f), (i)].
This behavior is, however, a result of different data analysis procedure:
The analytical model evaluates entanglement of the \emph{average state} obtained by postselection
    but the numerical simulations reveal the \emph{average entanglement} that we can recover.
In the extreme case of a perfect measurement (i.e., dynamics described by the stochastic master equation
    $\dd\rho = \Gamma\D[\sigma_z^1+\sigma_z^2]\rho\dt + \sqrt{\Gamma}\H[\sigma_z^1+\sigma_z^2]\rho\dW$)
    and success probability 100 \% (corresponding to discarding the measurement record),
    the average state is a mixture of all possible measurement outcomes,
\begin{align}
    \rho &= \frac{1}{4}\ket{00}\bra{00} + \frac{1}{2}\ket{\Psi_+}\bra{\Psi_+} + \frac{1}{4}\ket{11}\bra{11} \nonumber\\
    &= \frac{1}{4}\left(\begin{array}{cccc} 1&0&0&0 \\ 0&1&1&0 \\ 0&1&1&0 \\ 0&0&0&1 \end{array}\right),
\end{align}
    which is a separable state.
In simulations, on the other hand, the maximally entangled state $\ket{\Psi_+}$ is generated
    with 50 \% probability and the average entanglement is equal to 0.5 ebit.

\begin{figure}
	\centering
	\includegraphics[width=\linewidth]{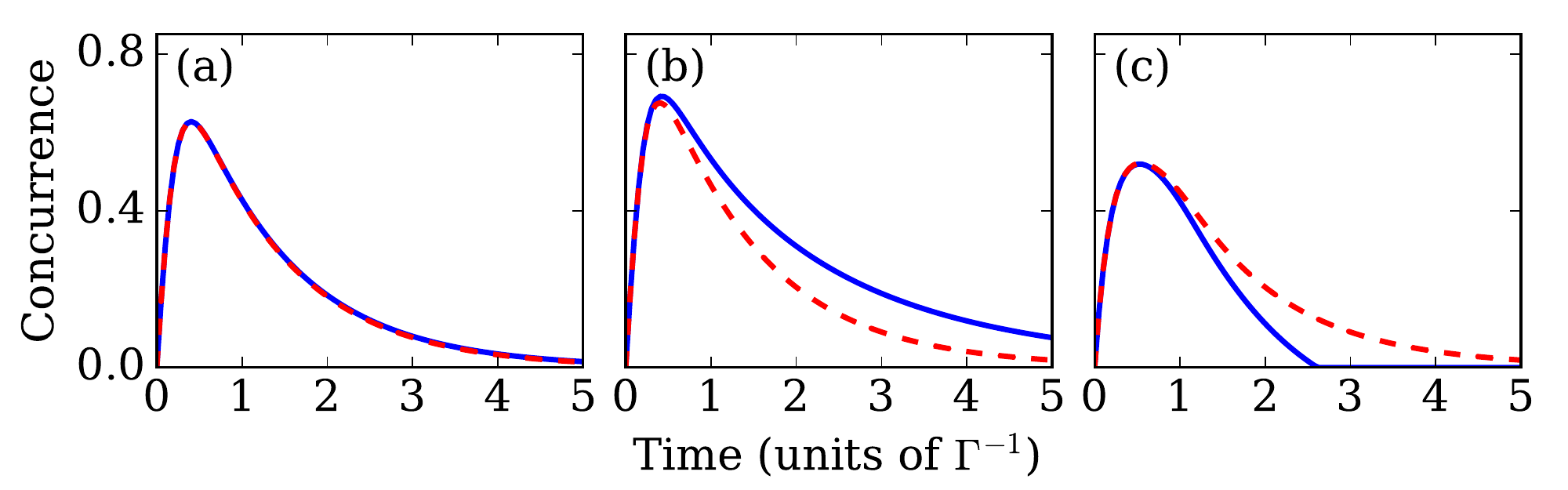}
	\caption{\label{fig.comparison}(Color online)
		Comparison of the total spin measurement (solid blue line) with the measurement of the spin difference (dashed red line).
		(a) Decoherence dominated by dephasing, $\gamma_1 = \gamma_2 = 0.2\Gamma$, $\gamma_- = 0.05\Gamma$, $P_\mathrm{succ} = 0.2$.
		(b), (c) Comparable dephasing and relaxation, $\gamma_1 = \gamma_2 = 0.1\Gamma$, $\gamma_- = 0.3\Gamma$,
            $P_\mathrm{succ} = 0.1$ (b), $P_\mathrm{succ} = 0.5$ (c).
		For all plots, the detection efficiency $\eta = 1$.}
\end{figure}

We can treat the measurement of spin difference $\sigma_z^1-\sigma_z^2$ similarly.
In this case, the entangled state $\ket{\Phi_+} = (\ket{00}+\ket{11})/\sqrt{2}$ can be generated
    from the initial state $\ket{\psi_0}$.
We compare this measurement with the total-spin measurement in Fig.~\ref{fig.comparison}.
When dephasing dominates the decoherence [panel (a)],
    the two strategies fare equally since dephasing affects both cases similarly.
When the qubit relaxation cannot be neglected, the success probability decides which measurement is preferable.
For small success probabilities [panel (b)], it is beneficial to measure the total spin $\sigma_z^1+\sigma_z^2$;
    with this measurement, only states in the relevant subspace (spanned by the states $\ket{01}$, $\ket{10}$) are postselected.
    Although similar statement is true also for the spin-difference measurement
    (where the relevant subspace is spanned by $\ket{00}$, $\ket{11}$), the ground state population contains also contributions
    from the decayed odd-parity states $\ket{01}$, $\ket{10}$ which reduce the concurrence.
The measurement of the spin difference is, however, a better choice if the success probability is large [panel (c)];
    in such a situation, the total-spin measurement mixes all three subspaces and the entanglement is reduced.
There is, nevertheless, a little difference between the two strategies for times up to the optimal measurement time,
    independent of the chosen success probability.

\section{Conclusions}\label{sec:conclusions}

We proposed a method to entangle superconducting qubits using optomechanical transducers
    and continuous homodyne measurements on the optical output.
We require only strong optomechanical cooperativity, $C = 4g^2/(\kappa\gamma\nbar)>\frac{1}{2}$,
    and sufficiently long qubit lifetimes; ground-state cooling of the mechanical oscillators or resolved-sideband regime are not necessary.
Owing to the topology of the scheme, where a single light beam travels through the optical cavities in a cascaded manner, 
    more nodes can be added to generate multipartite entanglement.
Although the presented setup works only with postselection, deterministic entanglement generation is possible with feedback
    \cite{Martin2015}.
    
The main advantage of the scheme, however, is its similarity to existing experiments.
Parity measurements are an established method for generation of entanglement in circuit QED \cite{Roch2014,Riste2013};
    optomechanical force sensing is a well-understood measurement strategy with a broad range of applications
    \cite{Rugar2004,Nichol2012,Moser2013,Schreppler2014,Clark2016,DeLepinay2016}.
Together, these techniques can extend quantum communication with superconducting circuits
    from a cryogenic environment to room temperature.
Their connection is, however, a difficult task---optomechanical force detection usually estimates classical forces;
    the question whether the quantum coherence of the source of such forces can survive during the measurement did not received enough attention.
By showing that thermal noise and optical losses play a surprisingly small role
    in measurement-induced generation of entanglement of superconducting qubits, we answered this question in the affirmative.

\begin{acknowledgments}
We thank Pertti Hakonen for useful discussions and Emil Zeuthen for critical reading of the manuscript.
This work was funded by the European Commission (FP7-Programme) through iQUOEMS (Grant Agreement No. 323924).
We acknowledge support by DFG through QUEST and by the cluster system team at the Leibniz University Hannover.
\end{acknowledgments}


%

\end{document}